# Preliminary Design of Scalable Hardware Integrated Platform for LLRF Application*

Lin Jiang, Jingjun Wen, Tao Xue[†], Xiaowei Guo, Haoyan Yang, Qiutong Pan,
Jianmin Li, Yinong Liu[1][2]
Liangjun Wei[3]
[1]Key Laboratory of Particle & Radiation Imaging (Tsinghua University), Ministry of Education
[2]Department of Engineering Physics, Tsinghua University, Beijing, China
[3]Greater Bay Area National Center of Technology Innovation, Guangzhou, China

## Abstract

In this paper, the SHIP4LLRF (Scalable Hardware Integrated Platform for LLRF) based on 6U VPX-standard was designed preliminarily, which includes 6U mother board and two HPC FPGA mezzanine cards (FMCs). The ADC and DAC FMC is based on ADS54J60 from TI and LTC2000Y-16 form ADI, respectively. The system mother board is based on Xilinx's Kintex UltraScale KU060, which also features 64-bit DDR4 SDRAM, QSFP and USB3.0 interfaces. Each FMC connector is assigned 58 pairs of LVDS standard IOs and 8 pairs of GTH high-speed serial lanes. Besides, the mother board is equipped with the self-developed ZYNQBee2 module based on ZYNQ7010 for slow control such as EPICS. All ADC or DAC raw data in each SHIP4LLEF is compressed lossless without triggering and transmitted to the process board. A scalar quantization method which is in development is used for lossless compression of ADC raw data, the process board will decompress the ADC data and perform a digital algorithm to measure the amplitude and phase of the high frequency signal. This design is scalable for testing and upgradability, meanwhile, the trigger-less data transmission enable this system participate in both local (rack-scale) and accelerator-wide communication networks.

## INTRODUCTION

For low-level RF (LLRF) systems in different frequency bands, their digital mainboards are generally versatile. Digital mainboards usually consist of ADC input, DAC output, EPICS communication, Acousto-Optic Modulator (AOM) driver, etc. Therefore, the design of the ADC and DAC parts will ultimately affect the amplitude stability and phase stability of the system, while control parts such as EPICS need to be simplified. At the same time, considering the demand for performance and flexibility of digital hardware as the scale of the LLRF system increases in the future, this article preliminarily designs a more dedicated platform for LLRF/BPM. The new platform prioritizes some advanced and demonstrated technologies and architectures.

In the future, we will implement more channels, higher readout bandwidth (up to 25 or 100 Gbps), higher sampling rate ADC/DAC (up to 1 GSPS), and a stronger LLRF digital hardware platform with synchronization between multiple channels. Providing a more flexible test and verification platform for ADC or DAC and RF subsystems will be introduced in detail in this paper.

## HARDWARE DESIGN

The hardware part consists of the system mother board and FMC mezzanine card, which will be introduced separately in the next section.

### ADC FMC Card

The FMC is based on the ADS54J60 from TI, which is 2-channel, 1000 MSPS, 16-bit JESD204B ADC. The power consumption is 1.35 W per ADC channel. 4 JESD204B lane output, lane rates up to 10 Gbps. DC input coupling is adopted to convert the input single-ended signal to differential signal by means of full-differential amplifier LMH6654 from TI. AD5686 and AD8676 from ADI are designed to bias the input signal to make full use of the full-scale dynamic range of ADS54J60, which is 1.9 V nominal. The actual object of FPGA mezzanine card (FMC) based on ADS54J60 is shown in Fig. 1. [1]

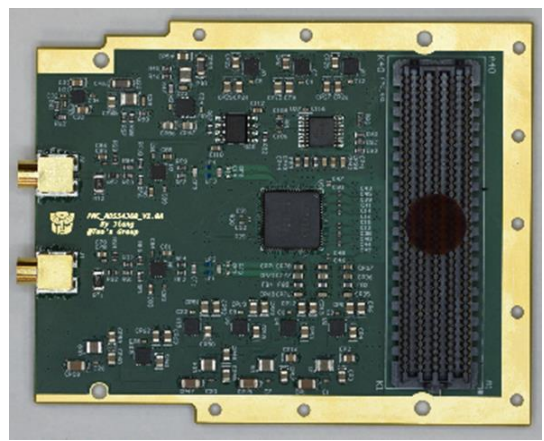

Figure 1: The actual object of FPGA mezzanine card (FMC) based on ADS54J60.

### 6U VPX-Standard Motherboard

The system mother board is a 6U VPX-standard module, as shown in Fig. 2. It combines two FMC high pin count (HPC) connectors with a Field Programmable Gate Array (FPGA). Each FMC connector is connected to the FPGA by 58 LVDS signals and 8 pairs of high-speed serial transceiver signals. The theoretical bandwidth of each lane is 16.3 Gbps. When the waveform digitizer works independently, the data can be read through USB3.0 interface or QSFP interface on the front panel. In addition, the trigger input/output, external reference clock input and pulse per second (PPS) signal input is also designed in front panel to ensure that a multi-channel (>4) readout electronic system can be built without the trigger and clock distribution board. A 64-bit DDR4 SDRAM is also designed to buffer the real tine data. ZYNQBee2 module self-developed based on Xilinx's ZYNQ7010 is used for slow control of the digitizer such as EPICS. Data in waveform digitizer is transmitted to the trigger board through eight GTH lanes through the backplane. It's important to note that the GTH lanes is re-driven to ensure signal integrity during backplane transmission.

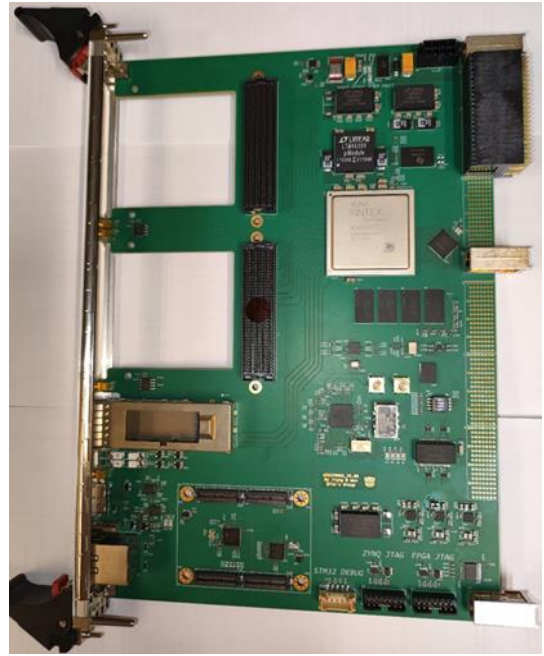

Figure 3: The actual object of 6U VPX-standard motherboard.

*FMC Mezzanine Card for LLRF*

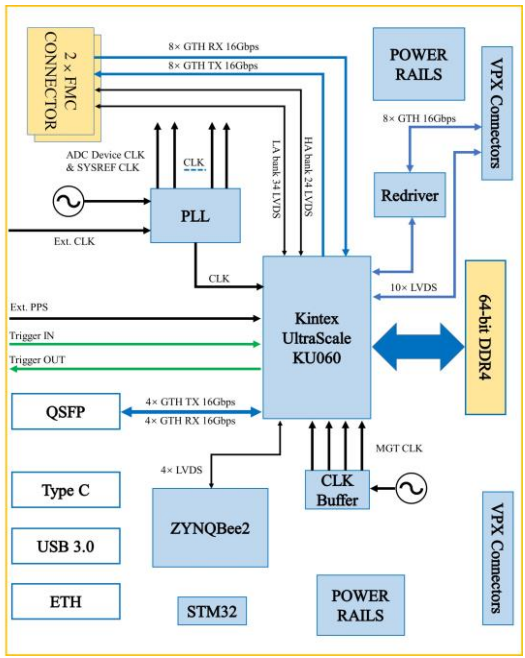

Figure 2: The schematic drawing of 6U VPX-standard digitizer.

The digitizer is a 16-layer PCB, hosting two FMC HPC connectors, one FPGA (Xilinx Kintex UltraScale KU060), one PLL (TI LMK04828), one DDR4 SDRAM, one redriver (TI DS280MB810), one USB3.0 module and other parts. In order to ensure the integrity of the power supply of the FPGA main electric rail, the via in pad (VIP) technology is used for design. The actual object of 6U VPX-standard digitizer is shown in Fig. 3.

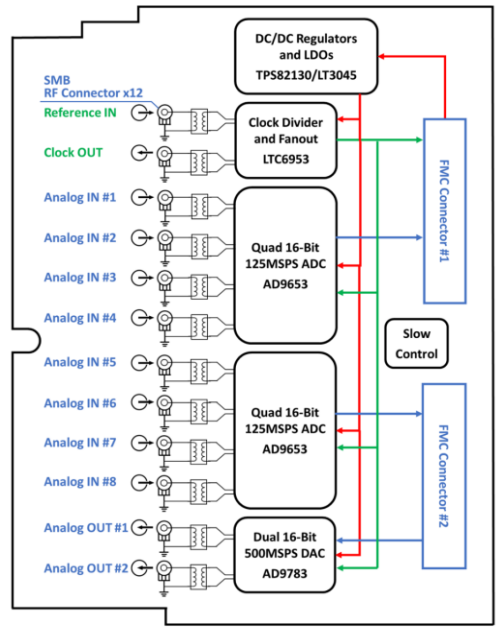

Figure 4: The schematic drawing of double-width FMC mezzanine card for LLRF.

At the same time, this paper aims to meet the needs of LLRF and preliminarily develops a double-width FMC mezzanine card. The ADC uses 4-chs 125 MSPS, 16-bit AD9653, the DAC uses 2-chs 500 MSPS, 16-bit AD9783, LTC6953 is used for clock distribution, and LT3045 ultralow noise, ultrahigh PSRR linear regulator is selected to

power the system. The schematic drawing of double-width FMC mezzanine card for LLRF is shown in Fig. 4. [2]

*FMC Mezzanine Card for BPM*

Another double-width FMC mezzanine card is designed for BPM requirements. It uses 4 2-chs, 250 MSPS, 16-bit ADS42JB69 from TI, and the power supply and clock distribution parts still use LTC6953 and LT3045 from ADI. The schematic drawing of double-width FMC mezza-nine card for BPM is shown in Fig. 5. [3]

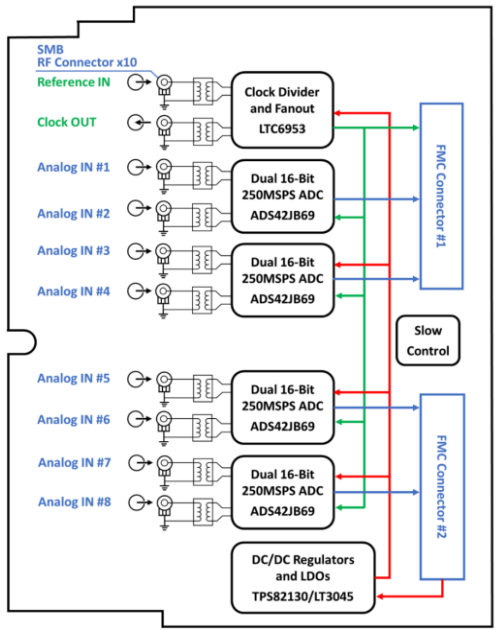

Figure 5: The schematic drawing of double-width FMC mezzanine card for BPM.

## HARDWARE TEST

We used a 100G loopback module to test the signal integrity of the QSFP interface on the front panel. In the test, the link speed was set to the theoretical 16.25 Gbps, the clock of the MGT bank was set to 156.25 MHz, and the encoding method was PRIBS-31. It is worth mentioning that the clock comes from a 156.25MHz crystal oscillator and LMK00334 clock buffer. The test results are shown in Table 1. The eye diagram opening area of each link is greater than 2600, and the opening UI is greater than 50%, and there are no errors under long-term transmission. The eye diagram test results of QSFP interface channel 1 is shown in Fig. 6.

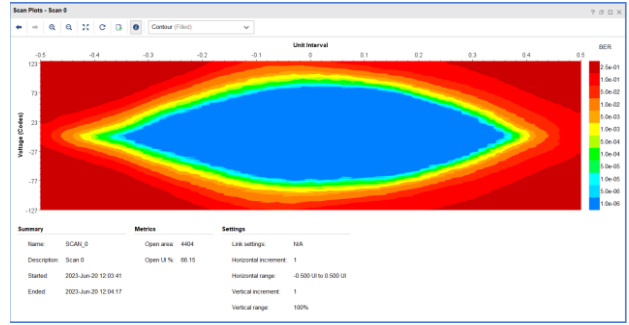

Figure 6: The eye diagram test results of QSFP interface channel 1.

Table 1: Front Panel QSFP Interface Signal Integrity Test

| MGT_B112 Channel | Open Area | Open UI |
| --- | --- | --- |
| Channel 1 | 4044 | 66.15% |
| Channel 2 | 3402 | 60.00% |
| Channel 3 | 2689 | 50.77% |
| Channel 4 | 3161 | 58.46% |

## CONCLUSION

This paper reports the preliminary design of FMC mezzanine card based on ADS54J60 and system motherboard based on 6U VPX-standard. Single block digitizer can support 4-channels 1000 MSPS, 16-bit sampling. In addition, the architecture diagrams of two different double-width FMC mezzanine cards designed for LLRF and BPM requirements are also introduced. In the actual test, the front panel QSFP interface link can operate at a theoretical speed of 16.25 Gbps without errors, which shows that data transmission and reception can be performed at a theoretical maximum of 64 Gbps. The hardware architecture studied in this article will provide an important foundation for subsequent ADC or DAC test and upgrades, channel number upgrades and data transmission bandwidth upgrades, and provide a reference for future hardware design applied to LLRF digital mainboards.

## ACKNOWLEDGEMENTS

We are grateful for the patient help of Yu Xue, Wenping Xue, and Jianfeng Zhang. They are seasoned full-stack hardware technologists with a wealth experience of in solder and rework in the electronics workshop at DEP.